\documentclass{article}

\pdfoutput=1

\usepackage{appendix}
\usepackage{algorithm}
\usepackage{algpseudocode}
\usepackage{arxiv}
\usepackage[utf8]{inputenc} 
\usepackage[T1]{fontenc}    
\usepackage{url}            
\usepackage{amsfonts}       
\usepackage{amsmath} 
\usepackage{amssymb}  
\usepackage{mathtools}
\usepackage{float}
\usepackage{accents}
\usepackage{subcaption}
\usepackage{graphicx}
\usepackage[usestackEOL]{stackengine}
\usepackage{multirow}

\DeclareMathOperator*{\argmin}{arg\,min}
\newcommand{\dfppartial}[2]{\frac{\partial \delta {#1}}{\partial {#2}}\Bigr|_{x_0}}

\title{A Distributed Model Predictive Wind Farm Controller for Active Power Control}

\author{
  \textbf{Valentijn van de Scheur}\\ 
  \texttt{} \\
  \textbf{Sjoerd Boersma}\\ 
  \texttt{} \\
}

\begin{document}

\maketitle

\begin{abstract}
	\vspace{10ex}
	Due to the fluctuating nature of the wind and the increasing use of wind energy as a power source, wind power will have an increasing negative influence on the stability of the power grid. In this paper, a model predictive control strategy is introduced that not only stabilizes the power produced by wind farms, but also creates the possibility to perform power reference tracking with wind farms. With power reference tracking, it is possible for grid operators to adapt the power production to a change in the power demand and to counteract fluctuations that are introduced by other power generators. In this way, wind farms can actually contribute to the stabilization of the power grid when this is necessary instead of negatively influencing it. A low-fidelity control-oriented wind farm model is developed and employed in the developed distributed model predictive controller. In this control model, the wake dynamics are taken into account and consequently, the model's order is relatively large. This makes it, from a computational point of view, challenging for a centralized model predictive control to provide real-time control for large wind farms. Therefore, the controller proposed in this paper is a distributed model predictive control. Here, the central control problem is divided into smaller local control problems that are solved in parallel on local controllers, which significantly reduces the computational complexity and brings the application of model predictive control in a wind farm a step closer to practical implementation. The proposed control solution is tested in simulations on a 10 and 64 turbine wind farm.
\end{abstract}
\clearpage

\section{Introduction}

With the increasing influence of greenhouse gases on the environment around the world, it is important to make the switch from fossil fuels to renewable energy sources. 
Fortunately, a lot of countries are on the way to become less dependent on fossil fuels~\cite{eurostat16}.
Among the renewable energy sources, wind energy is one of the largest contributors to the current power network~\cite{eurostat16} and it is expected to become an even larger contributor. Consequently, new challenges arise. As explained by the North American Electric Reliability Corporation (NERC)~\cite{GridBalancing}, the power demand on the electricity grid is constantly fluctuating. Power sources and loads should be able to counteract such fluctuations in order to keep the power line frequency constant. This is called frequency control. For most sources it is easy to provide frequency control services according to the NERC. However, as stated in~\cite{APCTutNREL14}, for wind turbines it is more difficult. The amount of power they supply is dependent on the wind speed. It is stated that, because of the fluctuating nature of the wind speed, the available power at wind turbines is constantly fluctuating. Therefore, it is not only difficult for wind turbines to provide frequency control services, but they even increase the instability on the grid~\cite{Vanfretti}. With the current share of electricity from wind turbines, this is not yet a major problem. However, with the increasing amount of wind turbines, this can become a serious issue that needs to be addressed~\cite{APCTutNREL14}. 

A solution is not only to stabilize the power output from wind turbines, but even letting wind farms provide frequency control services via control methods. Methods that strive to do this via controlling the power output, are called active power control (APC) methods. However, such APC architectures can be difficult to design. This is not only caused by the fluctuating nature of the available power at the turbines, but also because most wind turbines are placed together in so called wind farms. In these wind farms, wind turbines influence each other via their wakes. These are regions, which can be found downwind the turbines, having increased turbulence and wind speed deficit. These wakes make the dynamics within a wind farm complex and renders the APC for wind farms challenging. See the tutorial~\cite{WFTutorialBoersma} for a more detailed explanation on these challenges and a recent literature overview. 

An interesting control architecture for power reference tracking in wind farms is model predictive control (MPC). This type of control can take into account delayed dynamics, which are present within a wind farm. Indeed, these delays are mainly due to the wakes that travel from upstream to downstream turbines. As shown in~\cite{APCControlShapiro17,APC_Shapiro2018}, taking these dynamics into account can be beneficial for the tracking quality of the controller. Another interesting benefit of MPC is its predictive nature. Due to this, it is possible for a TSO to for see if a certain reference signal can be tracked sufficiently. This allows the TSO's to not only ensure a certain time-varying wind farm power injection in the grid, but also to push this power production to a limit, while taking turbine constraints into account. However, due to the centralized structure of these controllers, computational complexity can become an issue for large wind farms. Other centralized MPC controllers that try to circumvent this complexity by not taking wake dynamics into account in the controller have been presented in~\cite{APCControlBoersma18,APC_Sara2018} and a stochastic version in~\cite{APCBoersma2019_Stochastic}. 

In order to take both the computational complexity and the wake dynamics into account, it is interesting to use distributed model predictive control (DMPC) in wind farms so that a real-time application can become possible~\cite{SurveyWFControlKnudsen15}\footnote{In the context of this paper, real-time means that the time to update the control signals is smaller than the sample time of the controller, which in this work is 1 second.}. In DMPC, the optimization problem is separated over multiple controllers, thus requiring less communication and/or computational resources per controller. DMPC have been proposed in~\cite{DAPCControlH.Zhao16,DAPCControlV.Spudic15} though these controllers do not include a wake model. In~\cite{APCDistributed_Annoni2018}, a distributed controller that does take wake effects into account has been designed. This controller has been tested in a low-fidelity wind farm model and is focused on limiting the communication between the turbines. 

The contributions of this paper are
\begin{enumerate}
	\item the development of a controller wind farm model,
	\item the design of a DMPC for APC that can provide APC in real-time, while taking wake dynamics into account,
	\item a DMPC, which has a computational time that is nearly independent of the number of turbines in the farm,
	\item a DMPC that is validated in the medium-fidelity wind farm model WindFarmSimulator (WFSim).
\end{enumerate}
WFSim has been validated against flow and power data from a high-fidelity wind farm model~\cite{WFSim}. Indeed, the DMPC itself employs a controller wind farm model, which is developed in this paper. Clearly, the controller design in this work is focused on limiting the computational effort, such that it is possible to achieve real-time control.

\clearpage

\section{Controller Wind Farm Model}

This section develops the controller model used within the DMPC. This model consists of a flow (Section~\ref{subsection: Flow model}) and turbine model (Section~\ref{subsection:Turbine Model}). Both these models are combined (Section~\ref{subsection:combining models}) resulting in one (controller) state-space model (Section~\ref{subsection: State-Space Model}).

\subsection{Flow Model} \label{subsection: Flow model}

\begin{figure}[H]
	\centering
	\includegraphics[width=0.5\textwidth]{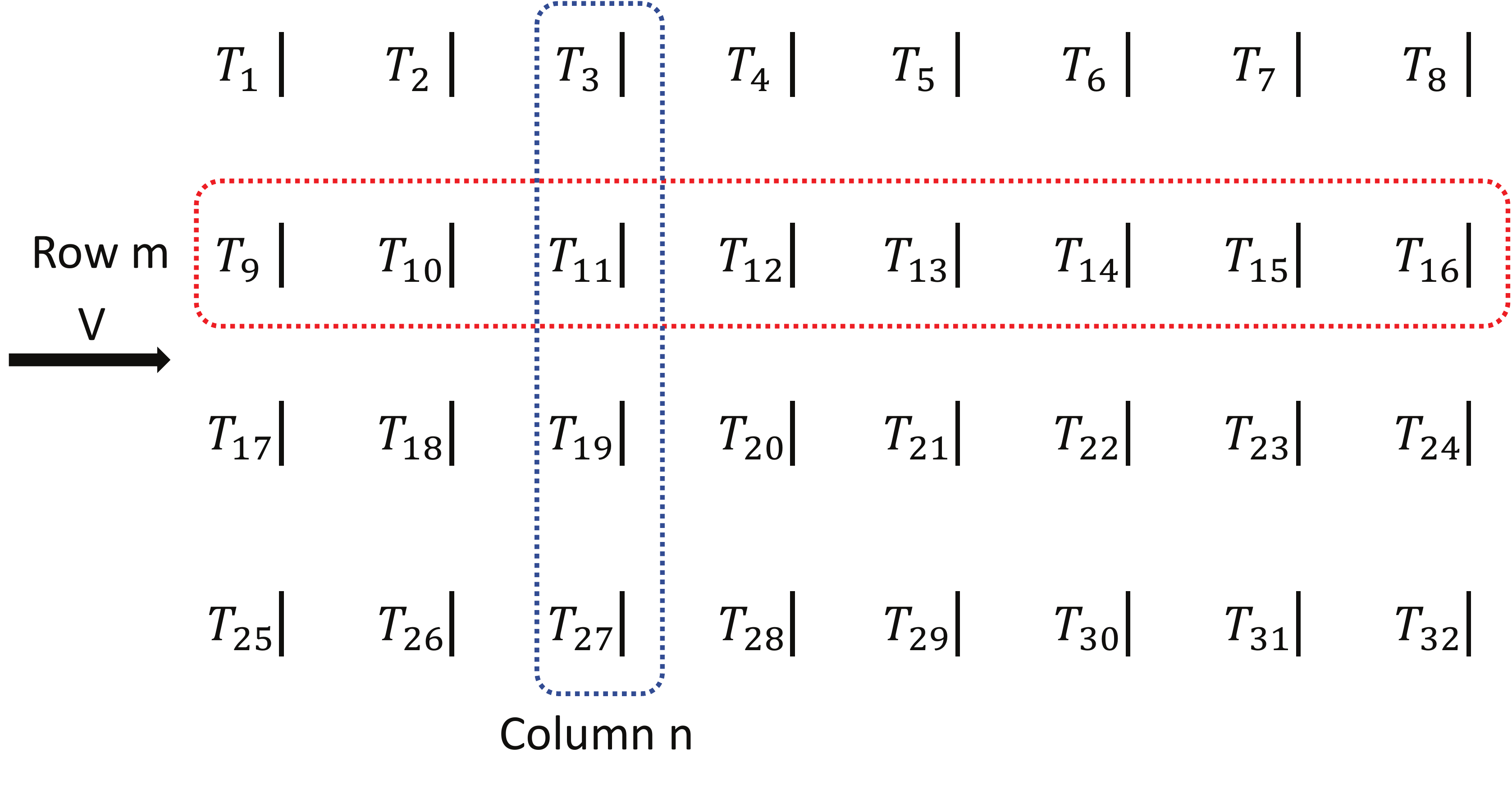}
	\caption{A wind farm is divided into rows $m$ and columns $n$. The wind turbines are depicted as the vertical black lines.}
	\label{fig:RowandColumns}
\end{figure}

The flow model is based on the model presented in~\cite{FrandsenModel06}. It is defined for rectangular wind farms consisting of $M$ parallel and equal rows and $N$ parallel and equal columns of wind turbines as shown in Fig.~\ref{fig:RowandColumns}. The wind turbines are depicted as the vertical black lines. The total number of wind turbines in the farm is denoted by $G=N\cdot M$. The rows are indexed with $m$, where $\{m\in\mathbb{Z}|1\leq m \leq M\}$, and the columns with $n$, where $\{n\in\mathbb{Z}|1\leq n \leq N\}$. The wind turbines $T_i$, $\{i\in \mathbb{Z}|1\leq i\leq G\}$, are numbered consecutively starting at the top row as shown in Fig.\ref{fig:RowandColumns}.  

Within the developed model, the interaction between the rows is ignored. Because of this, a separate model is defined for each row $m$. These separate models are combined to form a model for the complete wind farm. However, the model is defined for a single row $m$ firstly.

\begin{figure}[H]
	\centering
	\includegraphics[width=0.7\textwidth]{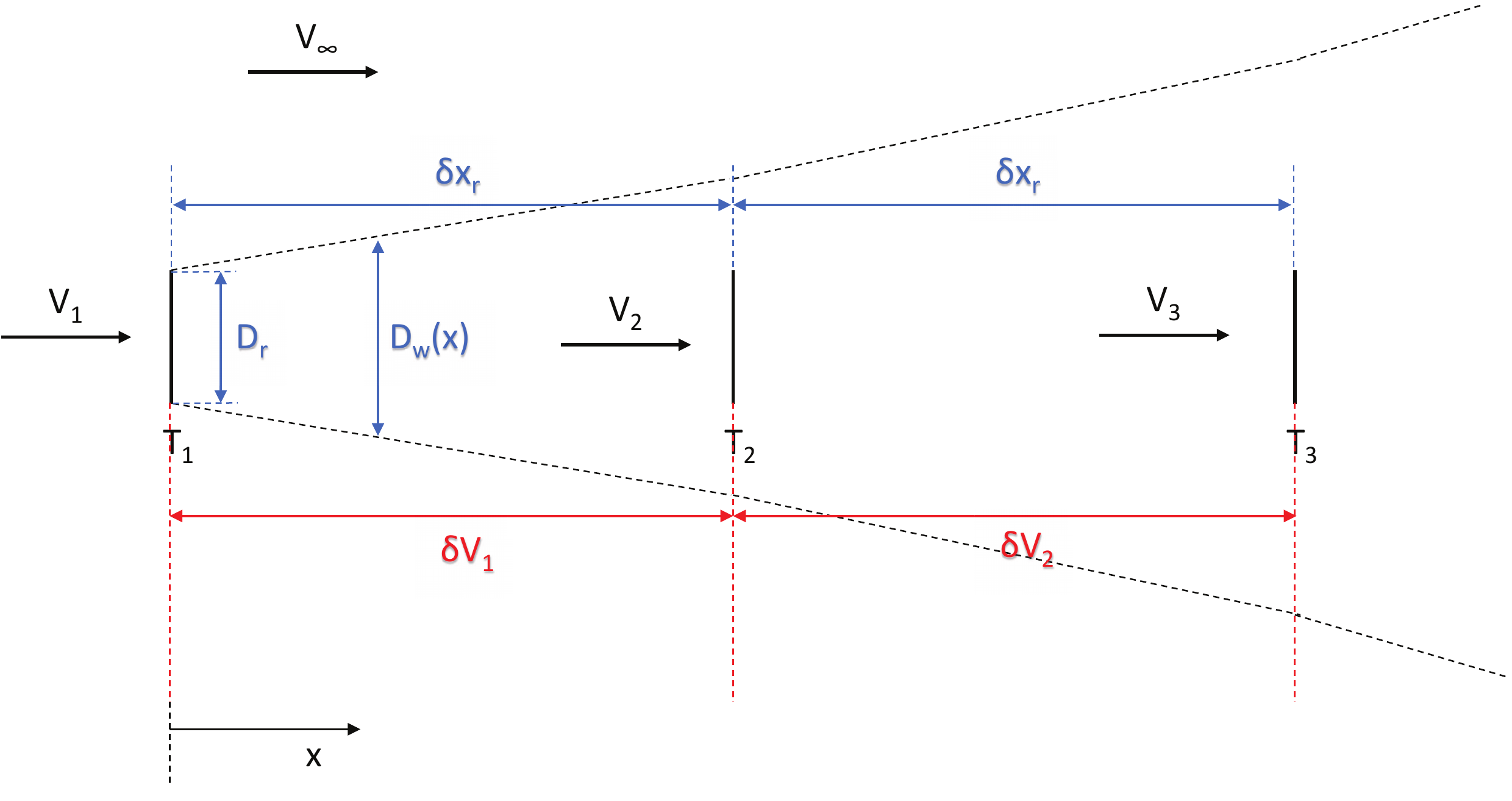}
	\caption{Close up on one row of wind turbines in fully waked conditions.}
	\label{fig:WindTurbineWakes}
\end{figure}

In Fig.~\ref{fig:WindTurbineWakes}, a close up of a part of one row is given. The wake zone is depicted as the area within the dotted black lines. 

The inflow wind speed into each wind turbine is given by $V_i$. This wind speed is defined as the wind speed just in front of the rotors. This is evaluated by $V_i={V_R}_i/(1-a_i)$, where ${V_R}_i$ is the inflow wind speed defined at the rotor of turbine $T_i$, and $a_i$ is the induction factor at turbine $T_i$. The wind speed far in front of the wind farm (free-stream wind speed) is defined by $V_\infty$. Furthermore, $\delta V_{i}$ is the wind speed deficit (the amount by which the wind speed is decreased in the wake) between turbine $T_i$ and $T_{i+1}$, {\it{i.e.}} $\delta V_i = V_{i}-V_{i+1}$. The diameter of the rotors is defined by $D_r$ and $\delta x_r$ is the distance between the turbines. As shown in Fig.~\ref{fig:WindTurbineWakes}, the wake zone diameter $D_w(x)$ gets wider over distance $x$ with a certain slope. This slope increases after each wind turbine depending on the local turbine setting, {\it{i.e.}} the thrust coefficients ${C_T}_i$.

The wind speed at the most upwind turbine, $V_1$, is taken equal to $V_\infty$ and is assumed to be known. The wind speed $V_i$ at each downwind turbine in the row is calculated as:

\begin{equation}
V_i=V_\infty-\sum^{i-1}_{j=1}\delta {V}_{j}.
\label{eq:DFPWindSpeedConStat}
\end{equation}

To simulate the time it takes for the wake to travel from wind turbine to wind turbine, the Taylor frozen turbulence hypothesis~\cite{Taylor'sFrozen38} is used. This hypothesis is approximated by

\begin{equation}
d_{j,i}=\text{round}\left(\frac{x_{j,i}}{V_\infty h}\right),
\end{equation}

where $d_{j,i}$ is the number of samples it takes for the wake to travel from turbine $T_j$ to $T_i$, $x_{j,i}$ is the distance between turbine $T_j$ and $T_i$ and $h$ is the sample time.
Using this, \eqref{eq:DFPWindSpeedConStat} is made time-dependent:

\begin{equation} \label{eq:difference equation}
V_i[k]=V_\infty-\sum^{i-1}_{j=1}\delta {V}_{j}[k-d_{j,i}],
\end{equation}

where the difference between time instance $k$ and $k+1$ is equal to sample time $h$.

Based on \cite{FrandsenModel06}, the wind speed deficit $\delta V_j[k]$ can be calculated as: 

\begin{equation}
\begin{aligned}
\delta V_j[k] &= \frac{1}{2}\frac{A_R}{A_{j+1}[k]}{C_T}_j[k]\left(V_j[k]-\frac{c_w}{1-c_w}\left(V_\infty[k]-V_j[k]\right)\right),
\end{aligned}
\label{eq:ViDiffReWr}
\end{equation}

where $c_w$ is a constant, $A_R$ is the rotor area, given by $A_R=\frac{\pi}{4}D_R^2$ and where $A_{j+1}[k]$ is the cross sectional area of the wake just in front of turbine $T_{j+1}$ calculated using the relation:

\begin{equation}
\begin{aligned}
A_{j+1}[k] & = A_R+\frac{1}{2}A_R\frac{c_w}{1-c_w}\sum^j_{l=1}{C_T}_l(k-d_{l,j}).
\label{eq:A_j}
\end{aligned}
\end{equation}.

This equation is made linear by differentiating it with respect to $V_j[k]$, ${C_T}_j[k]$ and $A_{j+1}$:

\begin{equation} \label{eq:linear wind speed difficit}
\begin{aligned}
\delta \hat{V}_j[k] &= \delta V_{j,0} + \dfppartial{V_j}{V_j}\Delta V_j[k] + \dfppartial{V_j}{{C_T}_j}\Delta {C_T}_j[k] + \dfppartial{V_j}{A_{j+1}}\Delta A_{j+1}[k] \\ 
&= \delta V_{j,0}+{\frac{1}{2}\frac{A_R}{A_{j+1,0}}{C_T}_{j,0}\left(1+\frac{c_w}{1-c_w}\right)\Delta V_j}[k]
+ {\frac{1}{2}\frac{A_R}{A_{j+1,0}}\left(V_{j,0}-\frac{c_w 
		2
	}{1-c_w}\left(V_{\infty}-V_{j,0}\right)\right)\Delta {C_T}_j}[k] \\ & \qquad - 
{\frac{1}{2}\frac{A_R}{A_{j+1,0}^2}{C_T}_{j,0}\left(V_{j,0}-\frac{c_w}{1-c_w}\left(V_{\infty}-V_{j,0}\right)\right) \Delta A_{j+1}}[k],
\end{aligned}
\end{equation}

where $x_0$ is the linearization point, $\delta \hat{V_j}[k]$ is the linearized version of $\delta V_j$, $\Delta V_j$ is the deviation from the linearization point $V_{j,0}$, {\it{i.e.}} $\Delta V_j[k]=V_j[k]-V_{j,0}$, $\Delta {C_T}_j[k]$ is the deviation from the linearization point ${C_T}_{j,0}$, {\it{i.e.}} $\Delta {C_T}_j[k] = {C_T}_j[k]-{C_T}_{j,0}$ and $\Delta A_{j+1}$ is the deviation from the linearization point $A_{j+1,0}$, which is evaluated as (using~\eqref{eq:A_j}):  

\begin{equation*}
\begin{aligned}
\Delta A_{j+1}[k]&=A_{j+1}-A_{j+1,0} =A_R+\frac{1}{2}A_R\frac{c_w}{1-c_w}\sum^j_{l=1}{C_T}_l(k-d_{l,j})-A_{j+1,0},
\end{aligned}
\end{equation*} 

in which $A_R$ is equal to $A_{1,0}$.

Now~\eqref{eq:linear wind speed difficit} and

\begin{equation}\label{eq:linear difference equation}
\hat{V_i}[k]=V_\infty-\sum^{i-1}_{j=1}\hat{\delta {V}}_{j}[k-d_{j,i}],
\end{equation}

together form the flow model for a single row of wind turbines.

\subsection{Turbine Model}\label{subsection:Turbine Model}

For modelling the turbines, the Actuator Disk Model~\cite{ADM_OG} is used.
The power of a wind turbine $T_i$ is then approximated by

\begin{equation}\label{eq:power function}
P_i[k] = \frac{1}{2}\rho V_i^3[k] A_R {C_T}_i[k] (1-a_i[k]),
\end{equation}

where $a_i[k]$ is the induction factor at turbine $T_i$ at time instant $k$. 

Differentiating~\eqref{eq:power function} with respect to $V_i[k]$ and ${C_T}_i[k]$ results in:

\begin{equation}\label{eq:linear power equation}
\begin{aligned}
\hat{P_i}[k] =&P_{i,0}+\dfppartial{P_i}{V_i}\Delta V_{i}[k] + \dfppartial{P_i}{{C_T}_i}\Delta {C_T}_i \\
= &P_{i,0}+\frac{3}{2}\rho V_{i,0}^2A_R {C_T}_{i,0}(1-a_{i,0}) \Delta V_{i}[k] + \frac{1}{2} \rho V_{i,0}^3A_R (1-a_{i,0}) \Delta {C_T}_i[k],
\end{aligned}
\end{equation}

in which $\hat{P_i}$ is a linearized $P_i$ and $P_{i,0}$ and $a_{i,0}$ are the linearization points of $P_i$ and $a_i$, respectively. The axial induction $a_i[k]=a_{i,0}$ in the above expression.

As shown, the input ${C_T}_i[k]$ influences the output $\hat{P}_i[k]$ directly, {\it{i.e.}} no turbine dynamics are taken into account. Following~\cite{MuntersFilter}, in order to add an approximation of the turbine dynamics and to smooth the input signal, a first order filter is added to the system: 

\begin{equation}
\tau \frac{d\tilde{C_T}_i[t]}{dt} + \tilde{C_T}_i[t] = {C_T}_i[t],
\end{equation}

in which $\tau$ is a time constant and $\tilde{C_T}_i[t]$ is the filtered version of ${C_T}_i[t]$. When discretized using a zero order hold, the filter is equal to

\begin{equation}\label{eq:discretized CT filter}
\begin{aligned}
\tilde{C_T}_i[k+1] 
&= e^{-\frac{1}{\tau}h}\tilde{C_T}_i[k] + \left(1 - e^{-\frac{1}{\tau}h}\right) {C_T}_i[k].
\end{aligned}
\end{equation}

This filter, the turbine model and the flow model are combined in the next subsection.

\subsection{Wind Farm Model}\label{subsection:combining models}

In this subsection, the flow model and turbine models are fused to create the wind farm model that is used in the developed controller. This model is referred to as the controller model. To account for mismatches between the model and reality, the tuning variables $c_{VV}, c_{VC_T}, c_{VA}, c_{PV}$ and $c_{PC_T}$ are included in the equations. Then, by combining the definitions for $\Delta V_i[k]$, $\Delta {C_T}_i[k]$ and $\Delta A_{i+1}[k]=A_R+\frac{1}{2}A_R\frac{c_w}{1-c_w}\sum^i_{l=1}{C_T}_l(k-d_{l,i})-A_{i+1,0}$
with~\eqref{eq:linear wind speed difficit}, \eqref{eq:linear difference equation}, \eqref{eq:linear power equation} and~\eqref{eq:discretized CT filter}, the fused model becomes:

\begin{equation}
\begin{aligned}
\hat{V_i}[k] &= V_{\infty,0}-\sum^{i-1}_{j=1}\delta \hat{V}_j[k-d_{j,i}]\\
\mathllap{\delta \hat{V}_j[k]} &= c_{VV}\dfppartial{V_j}{V_j}\hat{V_j}[k] + c_{VC_T}\dfppartial{V_j}{{C_T}_j}\tilde{C_T}_j[k]  + 
c_{VA}\dfppartial{V_j}{A_{j+1}}\frac{1}{2}A_R\frac{c_w}{1-c_w}\sum^j_{l=1}\tilde{C_T}_l(k-d_{l,j}) \\
&\qquad + {c_{\delta V}}_j \\
\hat{P_i}[k] &= c_{PC}\dfppartial{P_i}{V_i}\hat{V_{i}}[k] + c_{PC_T}\dfppartial{P_i}{{C_T}_i} \tilde{C_T}_i[k] + {c_{P}}_{i} \\
\tilde{C_T}_i[k+1] &= e^{-\frac{1}{\tau}h}\tilde{C_T}_i[k] + (1-e^{-\frac{1}{\tau}h}){C_T}_i[k],
\label{eq:CompleteSystem}
\end{aligned}
\end{equation}

where the partial fractions are defined above and the constant biases ${c_{\delta V}}_j$ and ${c_{P}}_i$ are given by:

\begin{align*}
{c_{\delta V}}_j &= \delta V_{j,0} - c_{VV}\dfppartial{V_j}{V_j}V_{j,0} - c_{VC_T}\dfppartial{V_j}{{C_T}_j}{C_T}_{j,0} + c_{VA}\left(\dfppartial{V_j}{A_{j+1}}A_{1,0} -  \dfppartial{V_j}{A_{j+1}}A_{j+1,0}\right)
\end{align*}
and
\begin{align*}
{c_{P}}_{i} &= P_{i,0} - c_{PV}\dfppartial{P_i}{V_i}V_{i,0} - c_{PC_T}\dfppartial{P_i}{{C_T}_i} {C_T}_{i,0}.
\end{align*}

The model for a single row of wind turbines within the farm is given by~\eqref{eq:CompleteSystem}.

To make the system valid for wind farms with an arbitrary number of rows, it is necessary to create such a system for each row $m$ within the farm. As the dynamics between the rows of turbines are not considered, these systems can just be stacked together to form a complete wind farm model. 

\subsection{State-Space Representation of the Wind Farm Model}\label{subsection: State-Space Model}

To use the model within the DMPC, it is written in state-space notation, which is easy to decompose in local subsystems following the notation as proposed in~\cite{DMPCBookLi2015}.

\begin{figure}[H]
	\centering
	\includegraphics[width=0.6\textwidth]{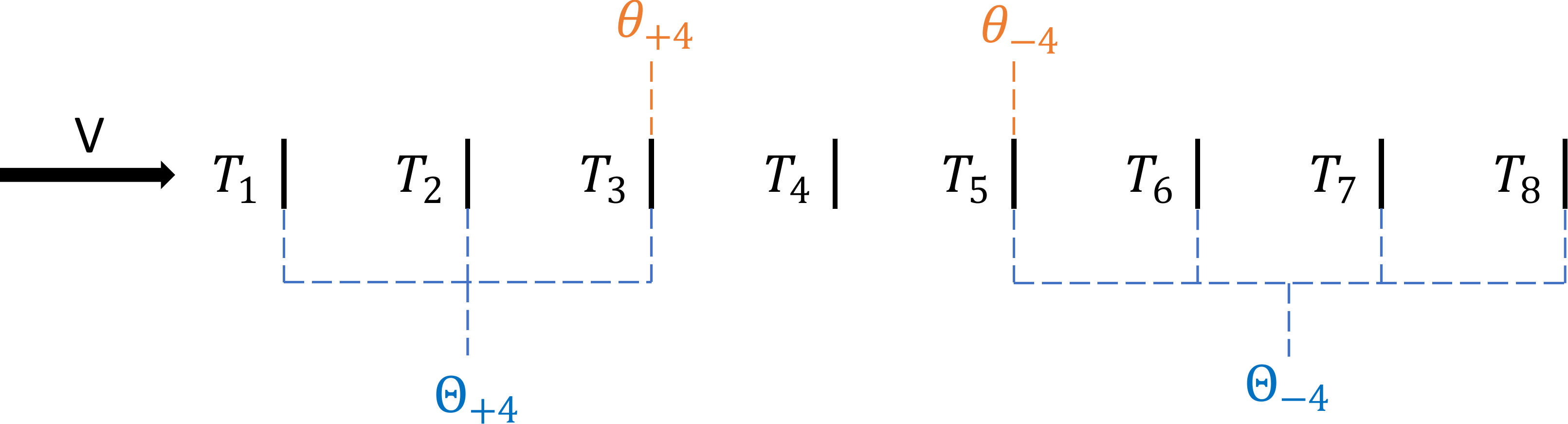}
	\caption{Visual representation of sets $\theta_{+4}$, $\theta_{-4}$, $\Theta_{+4}$ and $\Theta_{-4}$ in a 1x8 wind farm.}
	\label{fig:Thetatheta}
\end{figure}

The complete wind farm model developed in the previous section is denoted by $S$. If each wind turbine is taken as a subsystem, then $S$ consists of $G=N\cdot M$ subsystems $S_i$. Consequently, each local subsystem $S_i$ is denoted by

\begin{equation}
\label{eq:GenDecNot}
\begin{aligned}
x_i[k+1] &= A_{i,i}x_i[k]+B_{i}{C_T}_i[k]+\sum_{j\in\theta_{+i}}A_{i,j}x_j[k] + {c_x}_{i} \\
\hat{P}_i[k] &= C_{i}x_i[k] + {c_{P}}_{i},
\end{aligned}
\end{equation}

where ${c_{x}}_{i}$ and ${c_{P}}_{i}$ are constant biases for $S_i$, $x_i$ are the states of subsystem $S_i$, $A_{i,i}\in\mathbb{R}^{nx_i\times nx_i}$, $B_{i}\in\mathbb{R}^{nx_i\times 1}$ and $C_{i}\in\mathbb{R}^{1\times nx_i}$ are the system matrices of subsystem $S_i$ that connect the states $x_i[k]$ and inputs ${C_T}_i[k]$ to the states $x_i[k+1]$ and outputs $\hat{P}_i[k]$. $nx_i$ is the amount of states defined for subsystem $S_i$. $A_{i,j}\in\mathbb{R}^{nx_i\times nx_j}$ connects the states of subsystems $S_j\in \theta_{+i}$ to $S_i$, where $\theta_{+i}$ is the set of subsystems that directly influence $S_i$ through their states. Note that the subsystems only interact through their states. The states $x_i$ are defined in such a way, that only the subsystem directly upwind from and in the same row as $S_i$ is within $\theta_{+i}$. This does, however, not mean that the other subsystems upwind from subsystem $S_i$ have no influence on the states of $S_i$. The states of each subsystem in a row are influenced by the subsystem directly upwind from it. These effects are then, though the states, also passed down to the next downwind turbine. In this way, eventually, all the subsystems that are in the same row as $S_i$ and are upwind from $S_i$ will indirectly influence the states of $S_i$. The set of subsystems that indirectly and/or directly influence subsystem $S_i$ is denoted by $\Theta_{+i}$. This set, thus, consists of all subsystems that are in the same row as, and upwind from $S_i$. The set of subsystems that are indirectly and/or directly influence subsystem by $S_i$ is denoted by $\Theta_{-i}$. In Fig.~\ref{fig:Thetatheta} a visual representation of an example of the sets $\theta_{+i}$, $\theta_{-i}$, $\Theta_{+i}$ and $\Theta_{-i}$ is given for a wind farm consisting of 8 turbines in a single row.

\section{Control Architecture} \label{Chapter:Control Architecture}

The developed DMPC finds the optimal control sequence that steers the power output of a wind farm to a reference signal by solving a constrained optimization problem. To do so, it uses a controller model (introduced in the previous section) and measurements that are collected at time instance $k$ for initialization. After this, the cost function is minimized until time instance $k+H$, with $H$ the prediction horizon. Then, only the first instance of the found control sequence is applied. This sequence repeats itself every sample time, which is referred to as the receding horizon principle. 

In a centralized MPC architecture, measurements from all turbines in the farm are sent to a centralized controller. Then, all optimal control actions are determined at once and send to the turbines. Since a dynamical wind farm model is large in size, this solution is computationally heavy and consequently, real-time control is not feasible. Distributed controllers are a solution to this problem. In the proposed DMPC, the centralized control problem is divided into smaller sub-problems. These are solved in parallel by separate controllers, that share information with each other. In this way, it is possible to provide real-time control for large wind farms. 

\subsection{DMPC Integral Action} \label{Subsection: Velocity Form}

To ensure offset free tracking, integral action is added to the controller by transforming the decomposed controller model given in~\eqref{eq:GenDecNot} into the velocity form \cite{MPCVelocityTut,MPCServo}. By defining $\Delta x_i[k]=x_i[k]-x_i[k-1]$ and $\Delta {C_T}_i[k]={C_T}_i[k]-{C_T}_i[k-1]$ and introducing the new state ${x_I}_i[k] = \begin{bmatrix} \hat{P}_i^T[k] & \Delta x_i^T[k]\end{bmatrix}^T$, \eqref{eq:GenDecNot} is rewritten in velocity form:

\begin{equation}
\begin{aligned}
\underbrace{\begin{bmatrix}
	\hat{P}_i[k+1] \\
	\Delta x_i[k+1]
	\end{bmatrix}}_{{x_I}_i[k+1]}
=& 
\underbrace{\begin{bmatrix}
	I & C_{i}A_{i,i} \\
	0 & A_{i,i} \\
	\end{bmatrix}}_{{A_I}_{i,i}}
\underbrace{\begin{bmatrix}
	\hat{P}_i[k]\\
	\Delta x_i[k]
	\end{bmatrix}}_{{x_I}_i[k]}
+ 
\underbrace{\begin{bmatrix}
	0 & C_{i}A_{i,i-1} \\
	0 & A_{i,i-1} \\
	\end{bmatrix}}_{{A_I}_{i,i-1}}
\underbrace{\begin{bmatrix}
	\hat{P}_{i-1}[k]\\
	\Delta x_{i-1}[k]
	\end{bmatrix}}_{{x_I}_{i-1}[k]}
+
\underbrace{\begin{bmatrix}
	C_{i}B_{i} \\
	B_{i}
	\end{bmatrix}}_{{B_I}_{i}}
\Delta {C_T}_i[k] \\
\hat{P}_i[k] =& 
\underbrace{\begin{bmatrix}
	I & 0
	\end{bmatrix}}_{{C_I}_{i}}
\underbrace{\begin{bmatrix}
	\hat{P}_i[k]\\
	\Delta x_i[k] 
	\end{bmatrix}}_{{x_I}_i[k]}
\end{aligned}
\end{equation}

for all $\{i\in\mathbb{Z}|1\leq i\leq G,i\neq u_m~\forall m\}$, where $u_m$ is the index of the most upwind turbine in row $m$ and with the identity matrix $I\in\mathbb{R}^{1\times 1}$. For the subsystems $S_{u_m}$ in all rows $m$ the notation is the same, but without ${A_I}_{i,i-1}{x_I}_{i-1}[k]$, as these subsystems have no upwind subsystems that influence them.

\subsection{DMPC Matrix Development} \label{Subsection: Matrix Development}
In the DMPC, the controller model is propagated forward in time. This forward propagation of the presented state-space model results in extended matrices (directly used in the DMPC) that are presented now.

\begin{figure}[ht]
	\centering
	\includegraphics[width=0.6\textwidth]{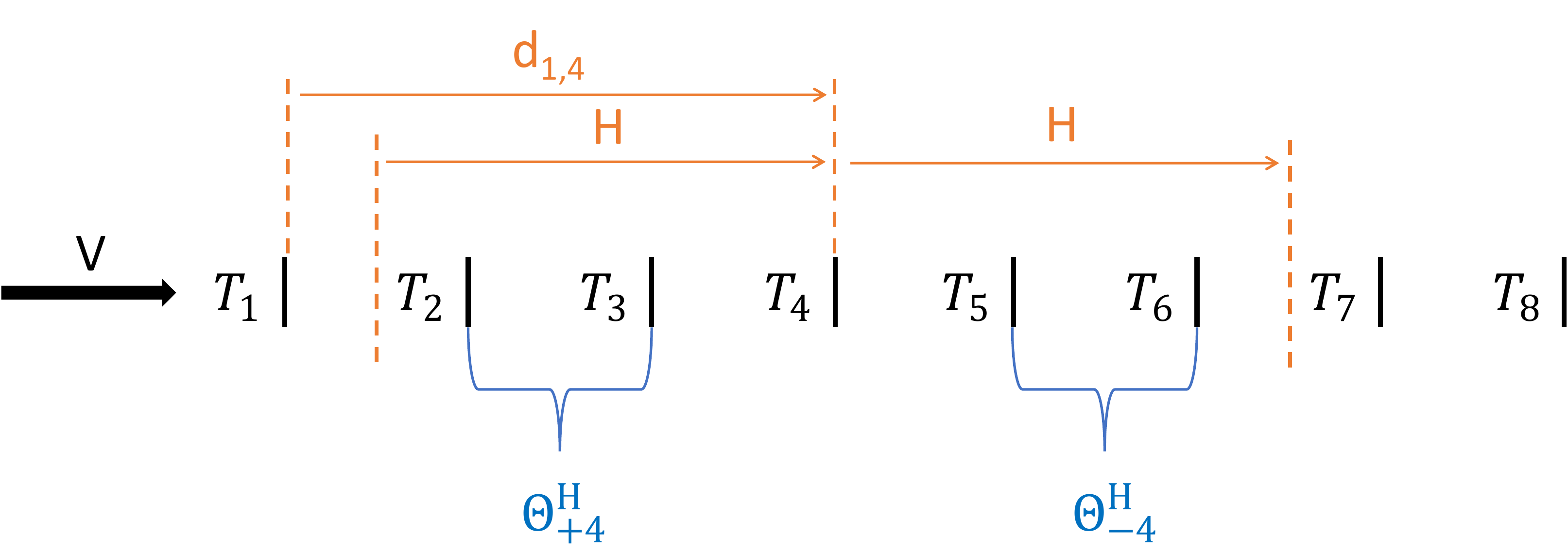}
	\caption{Visual representation of the sets $\Theta_{+4}^H=\{S_j|d_{j,4}\leq H\}$ and $\Theta_{-4}^H=\{S_j|d_{4,j}\leq H\}$ in an 1x8 wind farm in which $d_{1,4}>H$ and $H<d_{4,7}<d_{4,8}$.}
	\label{fig:ThetaH}
\end{figure}

By defining the vectors $Y_i=\begin{bmatrix}
\hat{P}_{i}[k+1] & \hat{P}_{i}[k+2] & \cdots & \hat{P}_{i}[k+H]
\end{bmatrix}^T$ for all $S_i$, where $H$ is the control horizon of the model predictive controller, it is possible to state that

\begin{equation}
\begin{aligned}
Y_i=&{F}_{i}x_{i}[k]+{\Phi}_{i}\Delta U_{i}+ \mathbf{C_I}_i\Xi_{i,i-1}\Bigg(\left(L_{i-1}{F_x}_{i-1}+W_{i-1}\right)x_{i-1}[k]+L_{i-1}{\Phi_x}_{i-1}\Delta U_{i-1} \\ & \quad +L_{i-1}\Xi_{i-1,i-2}\bigg(\left(L_{i-2}{F_x}_{i-2}+W_{i-2}\right)x_{i-2}[k]+L_{i-2}{\Phi_x}_{i-2}\Delta U_{i-2}+\dots \\ & \quad +L_{l+1}\Xi_{l+1,l}\Big(\left(L_{l}{F_x}_{l}+W_{l}\right)x_{l}[k]+L_{l}{\Phi_x}_{l}\Delta U_{l}\Big)\dots\bigg)\Bigg) \\
& \quad \forall\{i\in\mathbb{Z}|1 \leq i \leq G,i \neq u_m ~\forall m\},
\end{aligned}
\label{eq:YiDecRed}
\end{equation}

where $F_{i}=\mathbf{C_I}_{i}{F_x}_{i}$, with
${F_x}_{i}=
\begin{bmatrix}
{A_I}_{i,i} & {A_I}_{i,i}^2 & {A_I}_{i,i}^3 & \cdots & {A_I}_{i,i}^H
\end{bmatrix}^T$
and 
$$\mathbf{C_I}_{i}=
\begin{bmatrix}
{C_I}_{i} & 0 & \cdots & 0 \\
0 & {C_I}_{i} & \cdots & 0 \\
\vdots & \vdots & \ddots & \vdots \\
0 & 0 & \cdots & {C_I}_{i}
\end{bmatrix}.$$
Furthermore, $\Phi_{i}=\mathbf{C_I}_{i}{\Phi_x}_{i}$, with 
$$ {\Phi_x}_{i}=
\begin{bmatrix}
{B_I}_{i} & 0 & 0 & \cdots & 0 \\
{A_I}_{i,i}{B_I}_{i} & {B_I}_{i} & 0 & \cdots & 0 \\
{A_I}_{i,i}^2{B_I}_{i} & {A_I}_{i,i}{B_I}_{i} & {B_I}_{i}  & \cdots & 0 \\
\vdots & \vdots & \vdots & \ddots & \vdots \\
{A_I}_{i,i}^{H-1}{B_I}_{i} & {A_I}_{i,i}^{H-2}{B_I}_{i} &  {A_I}_{i,i}^{H-3}{B_I}_{i} & \cdots & {B_I}_{i}
\end{bmatrix}. $$ 
$$
\Xi_{i,i-1}= 
\begin{bmatrix}
{A_I}_{i,i-1} & 0 & 0 & \cdots & 0 \\
{A_I}_{i,i}{A_I}_{i,i-1} & {A_I}_{i,i-1} & 0 & \cdots & 0 \\
{A_I}_{i,i}^2{A_I}_{i,i-1} & {A_I}_{i,i}{A_I}_{i,i-1} & {A_I}_{i,i-1}  & \cdots & 0 \\
\vdots & \vdots & \vdots & \ddots & \vdots \\
{A_I}_{i,i}^{H-1}{A_I}_{i,i-1} & {A_I}_{i,i}^{H-2}{A_I}_{i,i-1} &  {A_I}_{i,i}^{H-3}{A_I}_{i,i-1} & \cdots & {A_I}_{i,i-1}
\end{bmatrix}$$ and $$
L_{i} = 
\begin{bmatrix} 
0 & 0 \\ I & 0 
\end{bmatrix}\in\mathbb{R}^{(nx_iH)\times (nx_iH)},
$$
with identity matrix $I\in\mathbb{R}^{\left(nx_i(H-i)\right)\times \left(nx_i(H-i)\right)}$ and $nx_i$ the number of states in $x_i$. Lastly,

\begin{equation*}
W_{i} = \begin{bmatrix} I \\ 0\end{bmatrix}\in\mathbb{R}^{(nx_iH)\times nx_i},
\end{equation*}

with identity matrix $I\in\mathbb{R}^{nx_i\times nx_i}$. $l$ in~\eqref{eq:YiDecRed} is the index of the most upwind turbine within the set $\Theta_{+i}^H$ which is a subset of $\Theta_{+i}$ containing the subsystems upwind from $S_i$ of which the wake effects can reach $S_i$ within the control horizon $H$. Remember that it takes time for wake effects to travel through the wind farm. The number of time samples it takes for wake effects to travel from turbine $T_j$ to $T_i$ was defined as $d_{j,i}$. Thus $\Theta_{+i}^H=\{S_j\in\Theta_{+i}|d_{j,i}\leq H\}$. It is not necessary to take the effects of subsystems $S_j$ that can not influence $S_i$ within $H$, thus $\{S_j\in\Theta_{+i}|d_{j,i}> H\}$, into account in the control problem of subsystems $S_i$. A visual representation of an example of a set $\Theta_{+i}^H$ is given in Fig.\ref{fig:ThetaH}. Here also the set of all turbines that is (in)directly influenced by $S_i$ within $H$, $\Theta_{-i}^H=\{S_j|d_{i,j}\leq H\}$, is depicted.

\subsection{DMPC Constraints} \label{Subsection: Constraints}
It is necessary to put constraints on the control actions. Theoretically, the maximum amount of energy is harvested from a flow of air by a single wind turbine when the thrust coefficient is at the Betz's limit~\cite{Betz'sLimit} (${C_T}_i=8/9$). If the wind turbine exceeds this Betz's limit, the rotor will rotate faster, but less energy is subtracted from the flow of air. Generally a wind turbine is not operated above this value. Also, if the controller model is linearized at a value below this Betz's limit, the model will predict increasing power outputs if the control actions increase above the Betz's limit. Therefore, it is chosen that ${C_T}_i$ should not exceed ${C_T}_{\rm{max}}=8/9$ for all $\{i\in\mathbb{Z}|1\leq i \leq G\}$.
Also, a requirement is that the wind turbines should not shut down completely. Therefore, it is decided that ${C_T}_i[k]$ can not be smaller than ${C_T}_{\rm{min}}=0.01$ for all $\{i\in\mathbb{Z}|1\leq i \leq G\}$.

Since the controller model is rewritten in the velocity form, the increments of the thrust coefficients, $\Delta {C_T}_i$, and not of the absolute values of the thrust coefficients, ${C_T}_i$, are used as control actions. Hence, it is not possible to put direct constraints on ${C_T}_i$. This is solved by realising that ${C_T}_i[k+j]={C_T}_i[k-1]+\sum^j_{l=0}\Delta {C_T}_i[k+l]$. Because at sample time $k$, ${C_T}_i[k-1]$ is known, it is possible to write the constraints for the complete prediction horizon as

\begin{equation}
\left.
\begin{array}{ccccc}
{C_T}_{\rm{min}} & \leq & {C_T}_i[k-1]+\Delta {C_T}_i[k]                 & \leq & {C_T}_{\rm{max}}\\
{C_T}_{\rm{min}} & \leq & {C_T}_i[k-1]+\Delta {C_T}_i[k]+\Delta {C_T}_i[k+1] & \leq & {C_T}_{\rm{max}} \\
&      & \vdots &      & \\
{C_T}_{\rm{min}} & \leq & {C_T}_i[k-1]+\sum^{H-1}_{j=0}\Delta {C_T}_i[k+j]      & \leq & {C_T}_{\rm{max}} \\  
\end{array}
\right\rbrace~\forall \{i \in \mathbb{Z} | 1 \leq i \leq G\},
\end{equation}

which is rewritten as

\begin{equation}
S_1\left({C_T}_{\rm{min}}-{C_T}_i[k-1]\right) \leq S_2 \Delta U_i[k] \leq S_1 \left({C_T}_{\rm{max}}- {C_T}_i[k-1]\right),
\end{equation}

where $\leq$ is here defined as an element-wise inequality and $S_1$ and $S_2$ are given by

\begin{equation}
\begin{aligned}
S_1 = \begin{bmatrix} 1 & \dotsm & 1 \end{bmatrix}^T \in \mathbb{R}^{H\times 1} \qquad \text{and} \qquad
S_2 = 
\begin{bmatrix}
1 & 0 & \dotsm & 0 & 0 \\
1 & 1 & \dotsm & 0 & 0 \\
\vdots & \vdots & \ddots & \vdots & \vdots \\
1 & 1 & \dotsm & 1 & 0 \\
1 & 1 & \dotsm & 1 & 1 \\
\end{bmatrix} \in \mathbb{R}^{H\times H}.
\end{aligned}
\end{equation}

$\Delta U_i[k]$ is a vector containing the control actions of subsystem $S_i$ for the complete time horizon $H$, thus $\Delta U_i[k]=\begin{bmatrix}\Delta{C_T}_i[k] & \Delta{C_T}_i[k+1] & \cdots & \Delta{C_T}_i[k+H-1]\end{bmatrix}^T$.

\subsection{DMPC Optimization Problem} \label{Subsection: Control Problem}

Using the constraints and model in velocity form, the control objective is given as

\begin{equation}
\begin{aligned}
&\min_{\Delta U_1,\cdots,\Delta U_G} \left \{ \left(\sum^G_{i=1}Y_i-Y_{\rm{ref}}\right)^TQ^TQ\left(\sum^G_{i=1}Y_i-Y_{\rm{ref}}\right) +\sum^G_{i=1}\Delta U_i^T R^T R \Delta U_i\right \}\\
& \text{s.t.} \quad S_1\left({C_T}_{\rm{min}}-{C_T}_i[k-1]\right) \leq S_2 \Delta U_i \leq S_1 \left({C_T}_{\rm{max}}- {C_T}_i[k-1]\right) \qquad \forall\{i\in\mathbb{Z}|1 \leq i \leq G\},
\end{aligned}
\label{eq:DisOptObj}
\end{equation}

with power reference $Y_{\rm{ref}}=\begin{bmatrix} P_{\rm{ref}}[k+1] & P_{\rm{ref}}[k+2] & \cdots & P_{\rm{ref}}[k+H]\end{bmatrix}^T$ and the weight matrices $Q=q\cdot I\in\mathbb{R}^{H\times H}$ and $R=r\cdot I\in\mathbb{R}^{H\times H}$ with $q>0$ and $r>0$. Note that this control problem is not coupled via the constraints, but only via the cost. This means that it is a so called coupled cost decoupled constraint (CCDC) problem, which is necessary for the control algorithm that is used in the developed controller. A shorter notation for the cost given in~\eqref{eq:DisOptObj} is given by:

\begin{equation}
\min_{\Delta U \in \bar{\underaccent{\bar}{\Delta U}}}f(\Delta U),
\end{equation}

where $\Delta U=\begin{bmatrix}\Delta U_1^T & \Delta U_2^T & \cdots & \Delta_G^T\end{bmatrix}^T$. The set $\bar{\underaccent{\bar}{\Delta U}}$ represents the constraints $S_1\left({C_T}_{\rm{min}}-{C_T}_i^T[k-1]\right) \leq S_2 \Delta U_i[k] \leq S_1 \left({C_T}_{\rm{max}}- {C_T}_i^T[k-1]\right)$ and $$f(\Delta U) =   \left(\sum^G_{i=1}Y_i-Y_{\rm{ref}}\right)^TQ^TQ\left(\sum^G_{i=1}Y_i-Y_{\rm{ref}}\right) +\sum^G_{i=1}\Delta U_i^T R^T R \Delta U_i.$$

\subsection{DMPC Algorithm}

In order to solve the above defined control problem, a control algorithm needs to be defined. The control algorithm used in this work is based on the Jacobi algorithm~\cite{Necoara2011,GPannocchiaJacobus} because it can solve CCDC problems in a distributed and parallel way~\cite{Necoara2011}.

In the Jacobi algorithm, at each iteration, local control problems $\mathbb{P}_z$ (with index $z$) are solved. In here, the most optimal control actions are sought not for all subsystems, but only for a single subsystem $S_i$ or a certain set consisting of multiple subsystems, while the control actions of all other subsystem are kept equal to the ones calculated at the previous iteration. Here, it is chosen to define these local control problems for sets consisting of a turbine $S_i$ and all the turbines within $\Theta_{-i}^H$. Notice that, if $H$ is kept equal, this set stays of equal size, regardless of the size of the wind farm. Also note that these sets of turbines will overlap each other. The solutions of these local control problems $\mathbb{P}_z$ are then combined using a weight.  

To this end, the transformation matrices $T_z$ and $\tilde{T}_z$ are defined such that

\begin{equation}
\Delta U = T_z^T\Delta U_z+ \tilde{T}_z^T\Delta \tilde{U}_z,
\end{equation}

where $\Delta \tilde{U}_z=\tilde{T}_z \Delta U$ contains the control actions of the subsystems that are kept equal to the values calculated at the previous iteration for local control problem $\mathbb{P}_z$. These control actions are, thus, kept constant while solving one local control problem. $\Delta U_z={T}_z \Delta U$ contains the control actions of the subsystems for which the optimal values are searched in local control problem $\mathbb{P}_z$. Thus, these control actions are manipulated by the algorithm to find the control actions for which the cost function is minimal. The Jacobi algorithm is defined in Algorithm~\ref{Alg: Jacobian}.

Here, local control problems are solved with this algorithm at every time step $k$, after which the first control actions of the optimal control sequences $\Delta U_i^{p}~~\forall\{i\in\mathbb{Z}|1\leq i \leq G\}$ are applied, the states $x_i[k]$ are measured and the procedure is repeated. 

By letting each subsystem calculate it's own $Y_i$ and sharing it with the other subsystems, the amount of required computational effort is kept minimal. Otherwise, each sub-problem would need to calculate the $Y_i$ of all the subsystems within the farm to get the value of the cost-function. Now, for a sub-problem $\mathbb{P}_z$, it is only necessary to calculate the $Y_i$'s of subsystems that are influenced within the time horizon $H$ by the subsystems considered in the local sub-problem $\mathbb{P}_z$.

Note that each sub-problem can be solved in parallel as indicated in Algorithm~\ref{Alg: Jacobian}. By doing so, the algorithm (hence controller) is equally fast regardless of the wind farm's size. As long as the number of subsystems taken into account in each sub-problem remains equal.
In the algorithm, the local optimization problems are solved using Gurobi 8.1.0 in MATLAB 2019a on a single 2.3 GHz Intel(R) Core(QM) with i7-3610QM processor. 
\clearpage
\begin{algorithm}
	\caption{Jacobian Algorithm that is used to solve the DMPC}
	\label{Alg: Jacobian}
	
	\begin{algorithmic}[1]
		
		\State Given $w_z > 0~~\forall \{z\in\mathbb{Z}|1\leq z\leq Z\},$ while $\sum w_z =1,\ p_{\rm{max}} > 0$, $\epsilon > 0$~~\text{and}~~$\Gamma > 1$
		\State Initialize $p = 0,\ \epsilon_i = \Gamma\epsilon,~~\forall z,\ \Delta U_{z}^0=0~~\forall z$
		\While{$\exists \epsilon_i > \epsilon~~\text{and}~~p\leq p_{\rm{max}}$}
		\ForAll{$\{z\in\mathbb{Z}|1 \leq z \leq Z\}$} \Comment{Do this in parallel}
		\State Solve the local control problem: \par \hskip\algorithmicindent $\Delta U_z^{p+1}= \argmin_{\Delta U_z \in \bar{\underaccent{\bar}{\Delta U}}_z}f(\Delta \tilde{U}_z^{p}, \Delta U_{z})$ 
		\State Construct "complete" local solution for $z$:\par
		\hskip\algorithmicindent $\Delta \hat{U}_z=T_z^T\Delta U_z^{p+1}+\tilde{T}_z^T\Delta \tilde{U}_z^{p}$
		\State Share $\Delta \hat{U}_z$
		\EndFor
		\State Combine the local solutions as $\Delta U^{p+1}=\sum_{z=1}^Z w_z \Delta \hat{U}_z$
		\State Share $\Delta U^{p+1}$
		\ForAll{$\{i\in\mathbb{Z}|1\leq i\leq G\}$} \Comment{Do this in parallel}
		\State $\epsilon_i = \left\lVert\Delta U^{p+1}_i - \Delta U^{p}_i\right\lVert$
		\State Calculate $Y_i$ using~\eqref{eq:YiDecRed}.
		\State Share $Y_i$ with all other subsystems.
		\EndFor
		\State $p = p+1$
		\EndWhile
		\State $\Delta U^{p}$ contains the optimal control sequence $~~\forall S_i\ \{i\in\mathbb{Z}|1\leq i\leq G\}$
	\end{algorithmic}
\end{algorithm}

\section{Simulation Results} \label{Chapter:ConTesting}

In this section the controller is tested on the medium-fidelity wind farm model WFSim~\cite{WFSim}. In this work, WFSim is regarded as the true wind farm. 

\begin{figure}[H]
	\centering
	\begin{subfigure}{.5\textwidth}
		\centering
		\includegraphics[width=.9\linewidth]{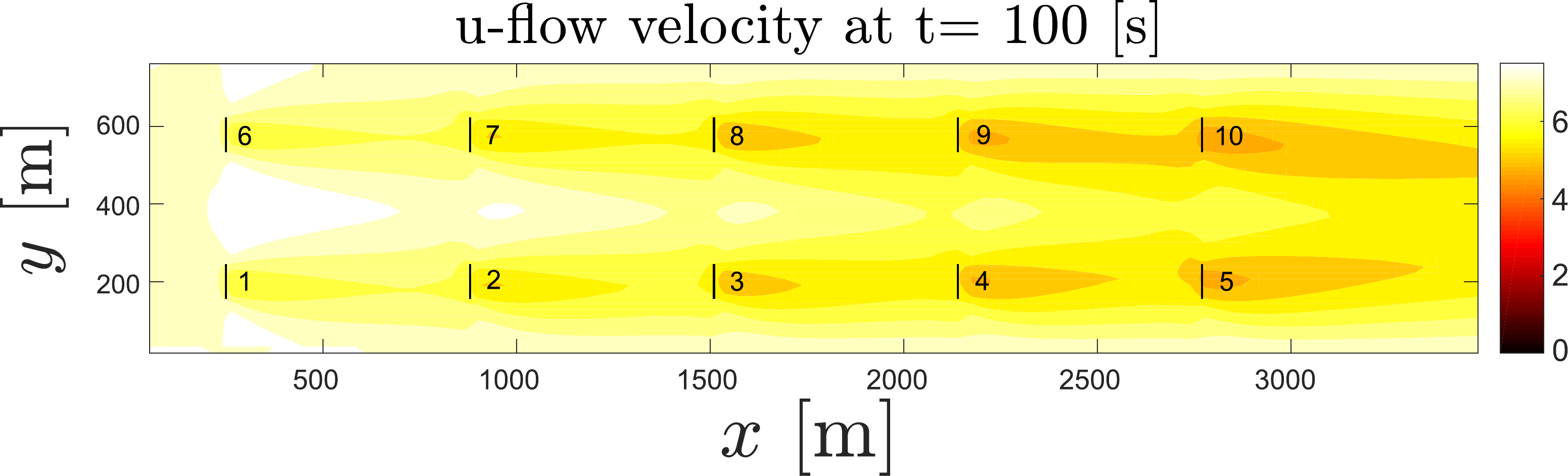}
		\caption{10 Turbine Wind Farm}
		\label{fig:VaalVsWFSim10T}
	\end{subfigure}%
	\begin{subfigure}{.5\textwidth}
		\centering
		\includegraphics[width=.9\linewidth]{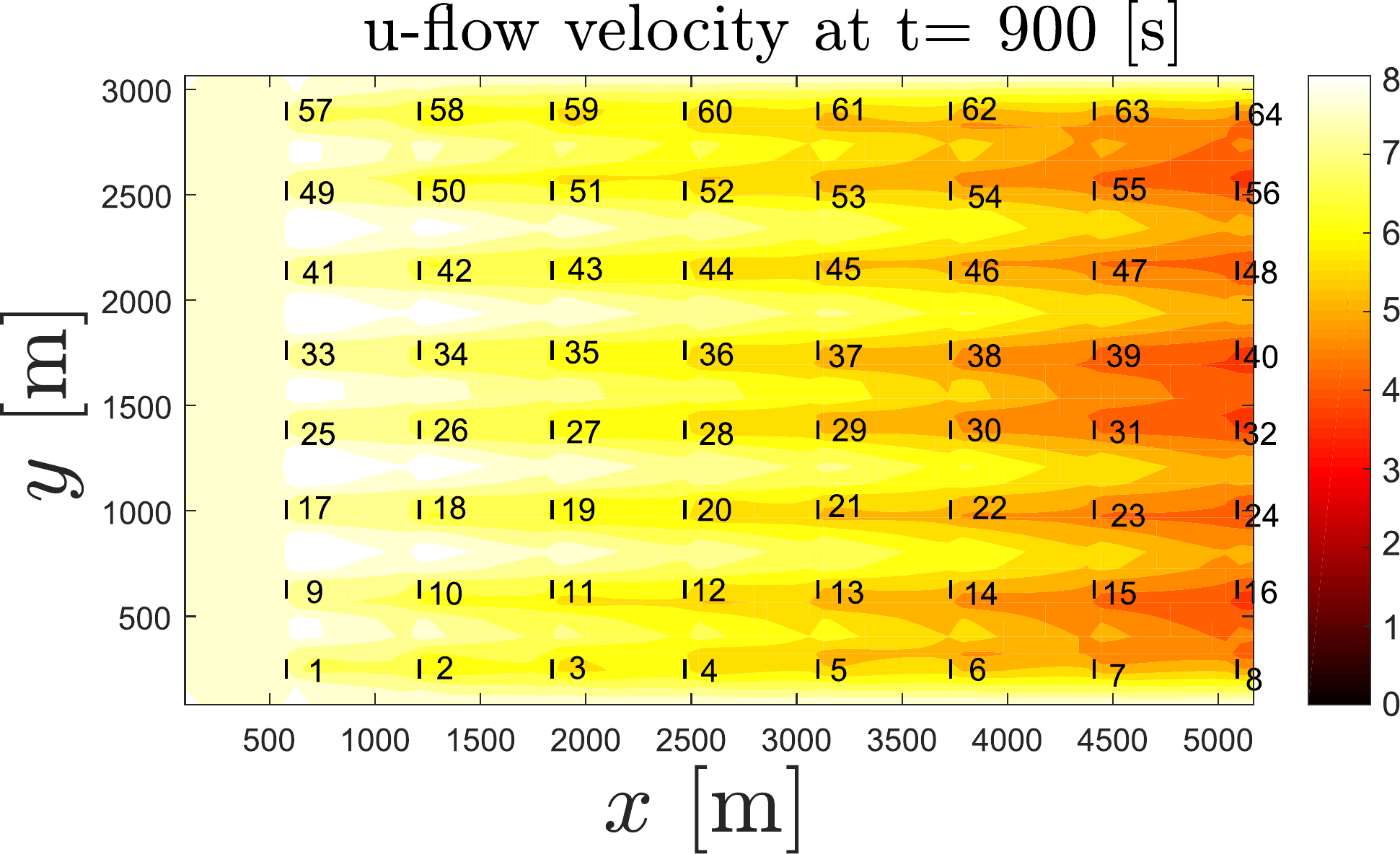}
		\caption{64 Turbine Wind Farm}
		\label{fig:64TCase}
	\end{subfigure}
	\caption{Visual flow representation of the 10 and 64 turbine wind farms in WFSim. The wind turbines are depicted by the vertical black lines and the flow is going from left to right with a free-stream wind speed of 7.5 m/s.}
	\label{fig:CasesWFSim}
\end{figure}

The controller will be tested in setups with $G=10$ and $G=64$ turbines under laminar inflow conditions. Both the 10 turbine and 64 turbine wind farm can be seen in Fig.~\ref{fig:CasesWFSim}. The distance between the turbines in both farms is 7 rotor diameters downstream and approximately 4 rotor diameters cross-stream. Other topology, WFSim settings and controller model parameters can be found in Table~\ref{Table: Setup Settings}. The controller model parameters are tuned such that this model is approximating WFSim power and flow behaviour. An detailed explanation of the WFSim parameters can be found in~\cite{WFSim}. 

The controller is tested with the reference signal
\begin{equation}
P_{\rm{ref}}[k] = \Big( 0.8 + \gamma \cdot \delta P[k] \Big) P_{\rm{greedy}},
\label{eq:Pref}
\end{equation}
where $P_{\rm{greedy}}$ is the power that the wind farm would produce if all its turbines would operate at the Betz limit. At this limit, the turbine will extract the theoretical maximum amount of energy from the airflow. This kind of control, where each turbine operates at its individual optimal settings, is called greedy control. Furthermore, $\delta P[k]$ is a normalized (its maximum value is one) RegD type AGC signal as defined in~\cite{PJMManual}. Notice that the reference signal $P_{\rm{ref}}[k]$ will exceed $P_{\rm{greedy}}$ for a certain period during the simulation if $\gamma>0.2$.

A measure for the tracking errors will be given by the root mean square error, which is defined as 
\begin{equation}
\text{RSME} = \sqrt{\frac{\sum^{N_s}_{k=1}(P_{\rm{ref}}[k]-{P}[k])^2}{N_s}},
\label{eq:RMSE}
\end{equation} 
where $N_s$ is the total number of time samples in the simulation.

As no computer with 64 cores is available to the authors of this paper, lines 4 until 8 and 11 until 15 in algorithm~\ref{Alg: Jacobian} will not be solved in parallel. However, it is still possible to give an indication for the time it would take to solve the controller problem if these lines were solved in parallel. This can be done by taking the maximum time that the algorithm spends on these lines for any of the sub-problems and adding this to the time it takes to run the rest of the controller.

\subsection{10 Turbine Wind Farm} \label{Section: 10 Turbine Case}

The 10 turbine farm consists of $M=2$ rows and $N=5$ columns of turbines. The chosen setup and controller settings can be found in Appendix~\ref{app:simulation settings}. For the prediction horizon, $H$, it is important to take the time delay in the wake effects between the turbines into account. For the 10T case the time delay between the most upwind turbine and the most downwind turbine is 304 seconds. Therefore, to include all the transients of the system, the time horizon should be larger than 304 seconds. However, for the controller this leads to a too computational heavy control problem to provide real-time control and therefore the controller is tested with $H=160$ seconds. As the time delay in the wake effects between three downwind turbines is 152 seconds, with this time horizon the wake effects between three downwind turbines are considered by the controller. It could also be argued that the effect of one turbine on the fourth downwind turbine is relatively small compared to the effect of that turbine on the first downwind turbines. In other words, the controller is still taking the most dominant effect into account when taking $H=160$ seconds.

\begin{figure}[ht]
	\centering
	\includegraphics[width=1\textwidth]{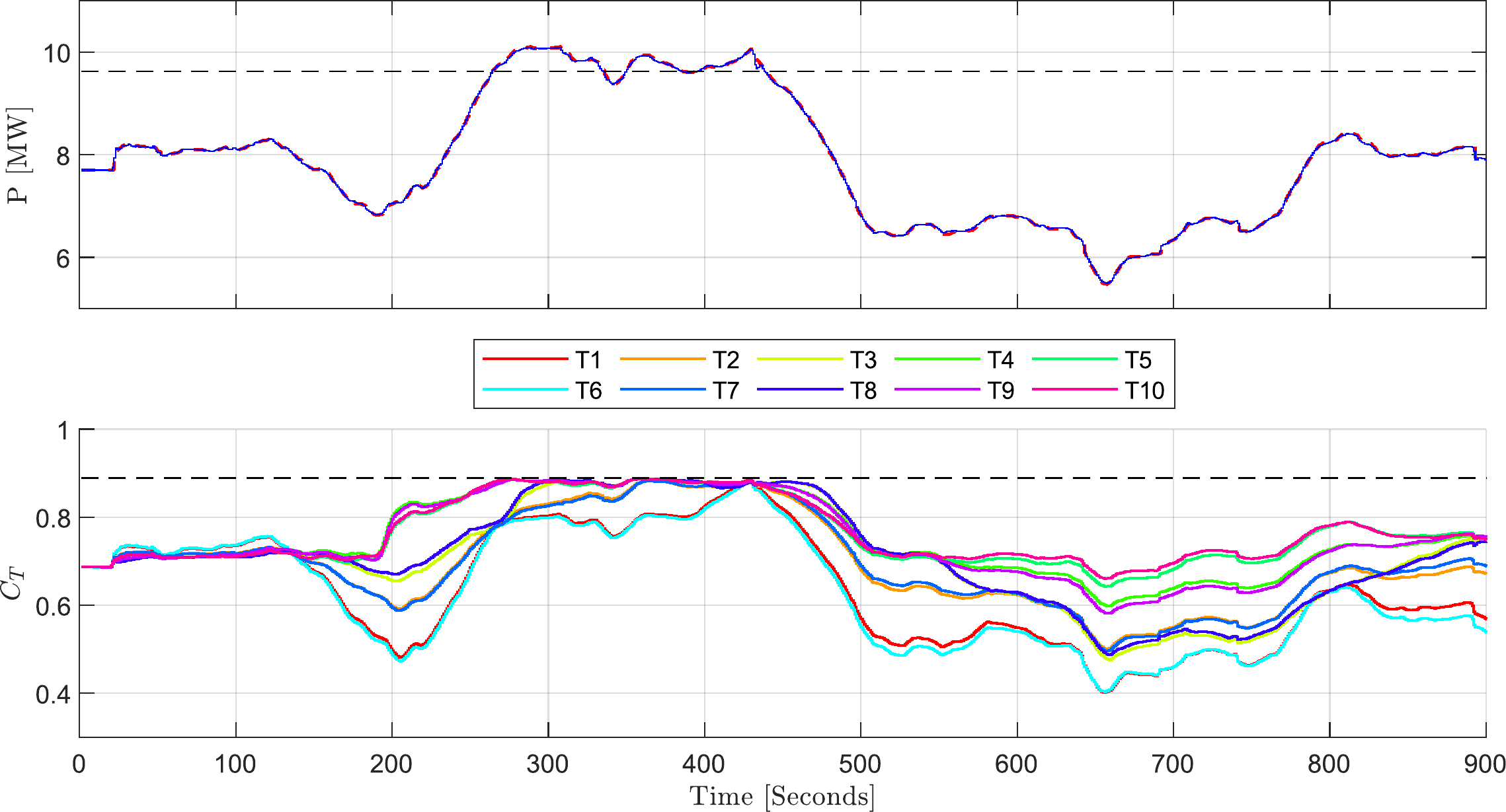}
	\caption{Tracking results for the 10 turbine wind farm with $\gamma=0.25$ (see~\eqref{eq:Pref}). 
		In the top figures $P_{\rm{ref}}[k]$ (red dashed), $\sum^G_{i=1}P_i$ (blue) and $P_{\rm{greedy}}$ (black dashed) can be seen. In the bottom figures ${C_T}_{\rm{max}}=8/9$ (dashed black) and ${C_T}_i\forall i$ can be seen.}
	\label{fig:Jac_norm_160_310_Mod_160_10T_0803}
\end{figure}

The simulation results are shown in Fig.~\ref{fig:Jac_norm_160_310_Mod_160_10T_0803}. As can be seen, the distributed controller is able to track a reference signal that exceeds $P_{\rm{greedy}}$ for a certain amount of time. In the control actions it is possible to see that the controller anticipates a surge in the reference. Before the reference exceeds $P_{\rm{greedy}}$, the controller decreases the $C_T$'s of the upwind turbines and increase the $C_T$'s of the downwind turbines. In this way, the wind speed in the wakes of the upwind turbines will increase. Because of the time delay in the wake effects, this higher wind speed will be available to the downwind turbines when $P_{\rm{greedy}}$ is exceeded by the reference. The RMSE$=0.019$ MW and the time it approximately takes to calculate the optimal control actions for a single time step if the algorithm is run in parallel is 0.92 seconds. This is below the defined threshold of 1 second.

\subsection{64 Turbine Wind Farm} \label{Section: 64 Turbine Case}

\begin{figure}[htb]
	\centering
	\includegraphics[width=1\textwidth]{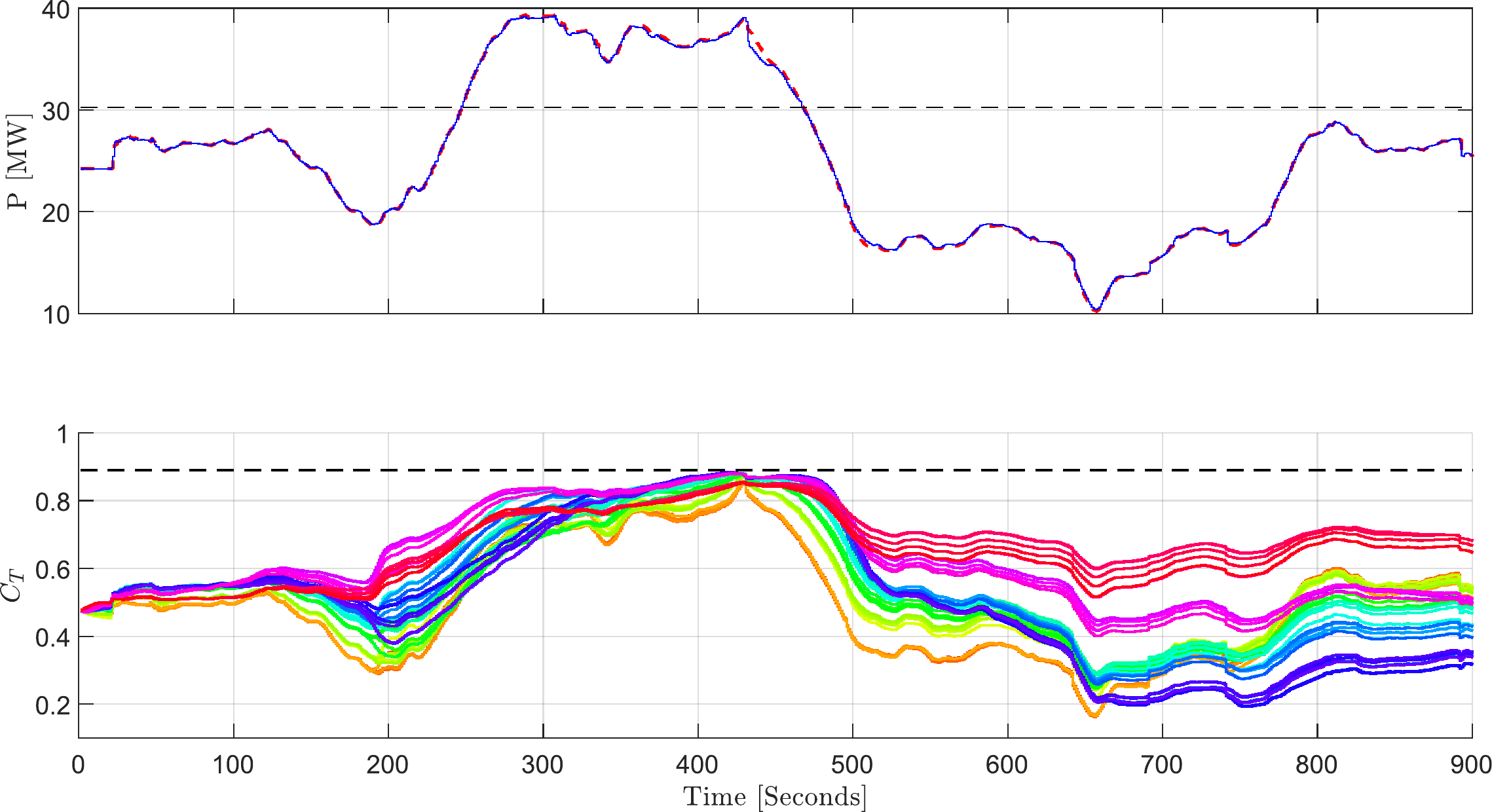}
	\caption{Tracking results for the 64 turbine wind farm with $\gamma=0.5$ (see~\eqref{eq:Pref}). 
		In the top figures $P_{\rm{ref}}[k]$ (red dashed), $\sum^G_{i=1}P_i$ (blue) and $P_{\rm{greedy}}$ (black dashed) can be seen. In the bottom figures ${C_T}_{\rm{max}}=8/9$ (dashed black) and ${C_T}_i\forall i$ can be seen.}
	\label{fig:Jac_norm_160_310_mod_160_64T_0802}
\end{figure}

For the 64 turbine wind farm, the controller is tested with $\gamma=0.5$ in the reference signal (see~\eqref{eq:Pref}) and prediction horizon $H=160$ seconds. Hence also here, wake effects between three downwind turbines are taken into account in the controller. Note that this reference signal significantly exceeds $P_{\rm{greedy}}$. The results can be seen in Fig.~\ref{fig:Jac_norm_160_310_mod_160_64T_0802}. As can be seen, the controller is able to track the reference signal properly with RMSE$=0.19$ MW. The time it approximately takes to calculate the optimal control actions for a single time step when the algorithm is running in parallel mode is 1.06 seconds, which is slightly above the threshold of 1 second. Notice that the computation time between the 10 turbine and 64 turbine wind farm is not exactly equal. This contradicts the idea that the controller would be equally fast, regardless of the size of the wind farm. The difference is, however, small and it is primarily caused by data handling in the controller. When solely looking at the time that the controllers would approximately spend on the lines 4 until 8 and 11 until 15 in Algorithm~\ref{Alg: Jacobian}, it can be concluded that these times are 0.91 seconds for the 10 turbine wind farm and 0.94 seconds for the 64 turbine wind farm. The difference between these times is considered negligible. Also in the 64 turbine wind farm it can be observed that the upwind turbines reduce their power production in order to pass through more wind for the downwind turbines. This results in good tracking behaviour of the controller. 

\clearpage

\section{Conclusions} \label{Chapter:Conclusions}

In order to ensure that the increase of wind energy penetration in the grid will continue, it is necessary to not only maximize its power output and minimize the fatigue loading of the turbines in the farm. It is also necessary to be able to provide grid services with a wind farm. One of these services in active power control. This paper proposed a computational efficient distributed model predictive controller that provides active power control (reference tracking), while its computational efficiency does not depend on the number of turbines in the farm.

From the simulation results it can be concluded that the developed controller is able to properly follow a time-varying reference signal, even if this exceeds $P_{\rm{greedy}}$. Although there are mismatches between the controller model and the real wind farm dynamics (WFSim in this work), the included integral action in the controller model causes the controller to compensate for these model mismatches such that tracking is still ensured. 

Furthermore, the developed controller is able to anticipate a surge in the reference signal that exceed $P_{\rm{greedy}}$ by decreasing $C_T$ for the upwind turbines and increasing $C_T$ for the downwind turbines. The controller is able to do this, because wake effects were taken into account in the controller model. As a consequence, the controller is able to ensure a power production more than $P_{\rm{greedy}}$ for a certain period of time.  

Also, the proposed distributed controller is able to provide real-time control in large wind farms. That is, if it is assumed that a processor core is available for each sub-problem. It has been shown that, under this assumption, the time it takes to calculate the control actions for a single time step stays approximately equal regardless of the size of the wind farm. More precisely, the controller takes around 1 second to calculate the control inputs for a single time step. This gives, however, little room for the communicational delays that were neglected in this paper. A solution to this would be to not update the control actions after each sample, but each two or three samples for example. Another solutions could be to increase the sample time and/or decrease the prediction horizon. In other words, the developed controller provides the flexibility to gain computational time if necessary.

Interesting to note is that besides creating the possibility to exceed $P_{\rm{greedy}}$ more and longer, taking the wake dynamics into account also makes it possible to provide additional information to the grid operators. With the control architecture proposed in this paper, it is possible to predict whether certain reference signals can or can not be tracked by a wind farm depending on the maximum available energy. This is useful information for grid operators and allows wind farms to provide better services.

Future work could focus on extending the controller model to work for wind farms with non-rectangular layouts and arbitrary wind directions. Also, to further validate the accuracy of the controller, it is necessary to test it in a high-fidelity wind farm model.

\clearpage
\appendix
\section{Setup and controller settings for simulations} \label{app:simulation settings}
In this appendix the setup and controller setting for the simulations can be found. The tuning parameters for the model and the weights $q$ and $r$ for the controller are found in an iterative manner. The time constant $\tau$ is chosen following W. Munters and J. Meyers \cite{MuntersFilter}.

\begin{table}[htb]
	\centering
	\caption{Simulation case details and settings for 10 turbine (10T) and 64 turbine (64T) case.}
	\begin{tabular}{c|c|c|c}
		& \multirow{2}{*}{Variable}   & \multicolumn{2}{c}{Value}  \\
		\cline{3-4}
		& & 10T Case & 64T Case \\
		\hline
		\multirow{7}{*}{\rotatebox{90}{Windfarm Setup}}
		& $G$ & 10 & 64 \\
		& $M$ & 5 & 8 \\ 
		& $N$ & 2 & 8 \\ 
		& $\delta x_r$ & 630 m & 630 m \\
		& $\delta y_r$ & 378 m & 378 m \\
		& $D_r$ & 90 m & 90 m \\
		& $V_{\infty}$ & 7.5 m/s & 7.5 m/s \\
		\hline
		\multirow{8}{*}{\rotatebox{90}{\Centerstack{Control \\ Model Settings}}}
		& $c_w$ & 0.68 & 0.31\\
		& $c_{VV}$ &  1.0 & 0.1\\
		& $c_{VC_T}$ & 1.0 & 0.6\\
		& $c_{VA}$ & 0.9 & 0.8\\
		& $c_{PV}$ & 1.0 & 0.9\\
		& $c_{PC_T}$ & 1.1 & 1.1 \\
		& $\tau$ & 5 & 5\\
		& $h$ & 1 & 1 \\
		\hline
		\multirow{6}{*}{\rotatebox{90}{\Centerstack{Control \\ Settings}}}
		& $q$ & 1 & 1\\
		& $r$ &  0.4 & 0.4\\
		& ${C_T}_{max}$ & 8/9 & 8/9 \\
		& ${C_T}_{min}$ & 0.01 & 0.01 \\
		& $p_{max}$ & 200 & 200\\
		& $\epsilon$ & $1\cdot10^{-2}$ & $1\cdot10^{-2}$\\
	\end{tabular} \qquad \quad
	\begin{tabular}{c|c|c|c}
		& \multirow{2}{*}{Variable}   & \multicolumn{2}{c}{Value}  \\
		\cline{3-4}
		& & 10T Case & 64T Case \\
		\hline
		\multirow{16}{*}{\rotatebox{90}{WFSim Settings}} &
		type & 'lin' & 'lin' \\
		& $L_x$ & 3500 & 5200\\
		& $L_y$ & 778 & 3146\\
		& $N_x$ & 100 & 80\\
		& $N_y$ & 50 & 40\\
		& powerscale & 1 & 1\\
		& forcescale & 1.25 & 1.25\\
		& $\mu$ & 0 & 0\\
		& $\rho$ & 1.2 & 1.2\\
		& $u_\infty$ & 7.5 & 7.5\\
		& $v_\infty$ & 0 & 0\\
		& $p_{init}$ & 0 & 0\\
		& lmu & 2 & 2\\
		& turbul & true & true \\
		& $n$ & 2 & 2\\
		& $m$ & 6 & 6 \\
		\multicolumn{4}{c}{}\\
		\multicolumn{4}{c}{}\\
		\multicolumn{4}{c}{}\\
		\multicolumn{4}{c}{}\\
		\multicolumn{4}{c}{}\\
	\end{tabular}
	\label{Table: Setup Settings}
\end{table}

\clearpage

\bibliographystyle{unsrt}

\begin{thebibliography}{10}
	
	\bibitem{eurostat16}
	Eurostat.
	\newblock Shares 2016 results, 2016.
	\newblock Data retrieved from Eurostat,
	\url{http://ec.europa.eu/eurostat/documents/38154/4956088/SHARES-2016-SUMMARY-RESULTS.xlsx}.
	
	\bibitem{GridBalancing}
	{NERC Resources Subcommittee}.
	\newblock Balancing and frequency control.
	\newblock Technical report, North American Electric Reliability Corporation,
	2011.
	
	\bibitem{APCTutNREL14}
	E.~Ela, V.~Gevorgian, P.~Fleming, Y.~C. Zhang, M.~Singh, E.~Muljadi,
	A.~Scholbrook, J.~Aho, A.~Buckspan, L.~Pao, V.~Singhvi, A.~Tuohy,
	P.~Pourbeik, D.~Brooks, and N.~Bhatt.
	\newblock Active power controls from wind power: Bridging the gaps.
	\newblock Technical report, National Renewable Energy Laboratory (NREL), 2014.
	
	\bibitem{Vanfretti}
	L.~Vanfretti, M.~Baudette, J-L. Domínguez-García, M.~S. Almas, A.~White, and
	J.~O. Gjerde.
	\newblock A phasor measurement unit based fast real-time oscillation detection
	application for monitoring wind-farm-to-grid sub–synchronous dynamics.
	\newblock {\em Electric Power Components and Systems}, 44(2):123--134, 2014.
	
	\bibitem{WFTutorialBoersma}
	S.~Boersma, B.~M. Doekemeijer, P.~M.~O. Gebraad, P.~A. Fleming, J.~Annoni,
	A.~K. Scholbrock, J.~A. Frederik, and Jan-Willem van Wingerden.
	\newblock A tutorial on control-oriented modeling and control of wind farms.
	\newblock In {\em American Control Conference}, 2017.
	
	\bibitem{APCControlShapiro17}
	C.~R. Shapiro, P.~Bauweraerts, J.~Meyers, C.~Meneveau, and D.~F. Gayme.
	\newblock Model-based receding horizon control of wind farms for secondary
	frequency regulation.
	\newblock {\em Wind Energy}, 20(7):1261--1275, 2017.
	
	\bibitem{APC_Shapiro2018}
	C.~R. Shapiro, J.~Meyers, C.~Meneveau, and D.~F. Gayme.
	\newblock Coordinated pitch and torque control of wind farms for power
	tracking.
	\newblock In {\em Annual American Control Conference}, 2018.
	
	\bibitem{APCControlBoersma18}
	S.~Boersma, B.~M. Doekemeijer, S.~Siniscalchi-Minna, and J-W. van Wingerden.
	\newblock A constrained wind farm controller providing secondary frequency
	regulation: {A}n {LES} study.
	\newblock {\em Renewable {E}nergy}, 134:639--652, 2019.
	
	\bibitem{APC_Sara2018}
	S.~Siniscalchi-Minna, F.~D. Bianchi, and C.~Ocampo-Martinez.
	\newblock Predictive control of wind farms based on lexicographic minimizers
	for power reserve maximization.
	\newblock In {\em American Control Conference}, 2018.
	
	\bibitem{APCBoersma2019_Stochastic}
	S.~Boersma, B.~M. Doekemeijer, T.~Keviczky, and J-W. van Wingerden.
	\newblock Stochastic model predictive control: uncertainty impact on wind farm
	power tracking.
	\newblock In {\em American Control Conference}, 2019.
	
	\bibitem{SurveyWFControlKnudsen15}
	T.~Knudsen, T.~Bak, and M.~Svenstrup.
	\newblock Survey of wind farm control-power and fatigue optimization.
	\newblock {\em Wind Energy}, 18:1333--1351, 2015.
	
	\bibitem{DAPCControlH.Zhao16}
	H.~Zhao, Q.~Wu, Q.~Guo, H.~Sun, and Y.~Xue.
	\newblock Optimal active power control of a wind farm equipped with energy
	storage system based on distributed model predictive control.
	\newblock {\em IET Generation, Transmission \& Distribution}, 10:669--677,
	2016.
	
	\bibitem{DAPCControlV.Spudic15}
	V.~Spudi{\'c}, C.~Conte, M.~Baoti{\'c}, and M.~Morari.
	\newblock Cooperative distributed model predictive control for wind farms.
	\newblock {\em Optimal Control Algorithms and Methods}, 36(3):333--352, 2015.
	
	\bibitem{APCDistributed_Annoni2018}
	C.~J. Bay, J.~Annoni, T.~Taylor, L.~Pao, and K.~Johnson.
	\newblock Active power control for wind farms using distributed model
	predictive control and nearest neighbor communication.
	\newblock In {\em American Control Conference}, 2018.
	
	\bibitem{WFSim}
	S.~Boersma, B.~M. Doekemeijer, M.~Vali, J.~Meyers, and J-W. van Wingerden.
	\newblock A control-oriented dynamic wind farm model: {WFS}im.
	\newblock {\em Wind Energy Science}, 3:75--95, 03 2018.
	
	\bibitem{FrandsenModel06}
	S.~Frandsen, R.~Barthelmie, S.~Pryor, O.~Rathmann, S.~Larsen, J.~H{\o}jstrup,
	and M.~Th{\o}gersen.
	\newblock Analytical modelling of wind speed deficit in large wind farms.
	\newblock {\em Wind Energy}, 9:39--53, 04 2006.
	
	\bibitem{Taylor'sFrozen38}
	G.I. Taylor.
	\newblock {The Spectrum of Turbulence}.
	\newblock {\em Proceedings of the Royal Society of London Series A},
	164:476--490, 02 1938.
	
	\bibitem{ADM_OG}
	W.~J.~M. Rankine.
	\newblock On the mechanical principles of the action of propellers.
	\newblock In {\em Transactions of the Institution of Naval Architects},
	volume~6, pages 13--30, 1865.
	
	\bibitem{MuntersFilter}
	W.~Munters and J.~Meyers.
	\newblock An optimal control framework for dynamic induction control of wind
	farms and their interaction with the atmospheric boundary layer.
	\newblock {\em Philosophical Transactions of the Royal Society A: Mathematical,
		Physical and Engineering Sciences}, 375(2091), 2017.
	
	\bibitem{DMPCBookLi2015}
	S.~Li and Y.~Zheng.
	\newblock {\em Distributed Model Predictive Control for Plant-Wide Systems}.
	\newblock John Wiley \& Sons Singapore Pte. Ltd., 2015.
	
	\bibitem{MPCVelocityTut}
	L.~Wang.
	\newblock A tutorial on model predictive control: Using a linear velocity-form
	model.
	\newblock {\em Developments in Chemical Engineering and Mineral Processing},
	12:573--614, 05 2008.
	
	\bibitem{MPCServo}
	M.~A. Stephens, C.~Manzie, and M.~C. Good.
	\newblock Model predictive control for reference tracking on an industrial
	machine tool servo drive.
	\newblock {\em IEEE Transactions on Industrial Informatics}, 9(2):808--816, May
	2013.
	
	\bibitem{Betz'sLimit}
	A.~Betz.
	\newblock {\em Introduction to the theory of flow machines}.
	\newblock Pergamon Press, 1966.
	
	\bibitem{Necoara2011}
	I.~Necoara, V.~Nedelcu, and I.~Dumitrache.
	\newblock Parallel and distributed optimization methods for estimation and
	control in networks.
	\newblock {\em Journal of Process Control}, 21(5):756--766, 2011.
	
	\bibitem{GPannocchiaJacobus}
	G.~Pannocchia, S.~J. Wright, B.~T. Stewart, and J.~B. Rawlings.
	\newblock Efficient cooperative distributed mpc using partial enumeration.
	\newblock {\em IFAC Proceedings Volumes}, 2009.
	
	\bibitem{PJMManual}
	PJM.
	\newblock Manual 12: Balancing operations.
	\newblock Technical report, PJM, 2018.
	
\end{thebibliography}


\end{document}